\documentclass[12pt,english]{article}
\usepackage[T1]{fontenc}
\usepackage[latin9]{inputenc}
\usepackage{geometry}
\usepackage{color}
\usepackage{babel}
\usepackage{amsmath}
\usepackage{amsthm}
\usepackage{graphicx}
\usepackage{setspace}
\usepackage[authoryear]{natbib}
\usepackage[unicode=true,pdfusetitle,
 bookmarks=true,bookmarksnumbered=false,bookmarksopen=false,
 breaklinks=false,pdfborder={0 0 1},backref=false,colorlinks=true]
 {hyperref}
\hypersetup{
 colorlinks=true,citecolor=green, urlcolor=cyan, linkcolor=black}

\makeatletter

\definecolor{green}{rgb}{0.05, 0.5, 0.06}

\makeatother

\begin{document}
\date{}
\title{\setstretch{1.5}{\huge{}Weak instruments in multivariable Mendelian randomization:
methods and practice}\thanks{Ashish Patel (\protect\href{mailto:ashish.patel@mrc-bsu.cam.ac.uk}{ashish.patel@mrc-bsu.cam.ac.uk});
James Lane (\protect\href{mailto:james.lane@mrc-bsu.cam.ac.uk}{james.lane@mrc-bsu.cam.ac.uk});\protect \\
Stephen Burgess (\protect\href{http://sb452@medschl.cam.ac.uk}{sb452@medschl.cam.ac.uk}).}}
\author{Ashish Patel{\small{} }\textsuperscript{{\small{}a}}, James Lane
{\small{}}\textsuperscript{{\small{}a}}{\small{},} \& Stephen Burgess{\small{}
}\textsuperscript{{\small{}a,b}}}
\maketitle
\begin{center}
{\small{}\vskip -2em}\textsuperscript{{\small{}a}}{\small{} MRC
Biostatistics Unit, University of Cambridge}\\
{\small{}}\textsuperscript{{\small{}b}}{\small{} Cardiovascular
Epidemiology Unit, University of Cambridge}\\
{\small{}~}{\small\par}
\par\end{center}
\begin{abstract}
The method of multivariable Mendelian randomization uses genetic variants
to instrument multiple exposures, to estimate the effect that a given
exposure has on an outcome \textsl{conditional} on all other exposures
included in a linear model. Unfortunately, the inclusion of every
additional exposure makes a weak instruments problem more likely,
because we require conditionally strong genetic predictors of each
exposure. This issue is well appreciated in practice, with different
versions of F-statistics routinely reported as measures of instument
strength. Less transparently, however, these F-statistics are sometimes
used to guide instrument selection, and even to decide whether to
report empirical results. Rather than discarding findings with low
F-statistics, weak instrument-robust methods can provide valid inference
under weak instruments. For multivariable Mendelian randomization
with two-sample summary data, we encourage use of the inference strategy
of \citet{Andrews2018} that reports both robust and non-robust confidence
sets, along with a statistic that measures how reliable the non-robust
confidence set is in terms of coverage. We also propose a novel adjusted-Kleibergen
statistic that corrects for overdispersion heterogeneity in genetic
associations with the outcome. 
\end{abstract}

\section{Introduction}

Mendelian randomization uses genetic variants as instrumental variables
to estimate the effect that an exposure has on an outcome in the presence
of unobserved confounding \citep{Smith2003}. Multivariable Mendelian
randomization \citep{Burgess2015} generalises this idea by using
genetic variants to instrument \textsl{multiple} exposures, to estimate
the effect that a given exposure has on an outcome adjusted for all
other exposures included in the model. This is a powerful approach
in that it offers a way to disentangle truly relevant exposures from
those that are only associated with the outcome through their association
with a causal exposure. 

In recent years, workhorse multivariable Mendelian randomization methods
have been creatively applied to explore a variety of interesting problems.
This includes studies of time-varying exposure effects \citep{Sanderson2022,Tian2024},
disease progression \citep{Lawton2024}, mediation \citep{Carter2021},
and causal mechanisms of pharmacological interventions \citep{Burgess2023}.
A common theme in many of these applications is \textsl{weak instruments},
and since there is an increasing interest in investigating causal
effect heterogeneity across individuals and time, it seems likely
that weak instrument issues will remain salient in multivariable Mendelian
randomization studies moving forward.

What do we mean by weak instruments? In Mendelian randomization studies
with one exposure, the concept of weak instruments is obvious: genetic
variants are only weakly associated with the exposure, if at all.
This can result in biased inferences on the exposure effect. In the
two-sample setting, where genetic associations with the exposure are
measured from a non-overlapping sample from genetic associations with
the outcome, exposure effect estimation using conventional methods
has been shown to be biased towards zero \citep{Pierce2013}. 

In multivariable Mendelian randomization, the concept of weak instruments
is more nuanced, both in terms of how we measure weak instruments,
as well as their consequence for inference. We require \textsl{conditionally}
strong genetic predictors of each exposure included in the model;
that is, strong genetic associations with the exposures that are not
collinear. In typical multivariable Mendelian randomization applications,
particularly in studies of drug target perturbation where only single
gene regions are considered for instrument selection, the challenge
is not so much in finding genetic variants that associate strongly
with several correlated exposures, but in finding genetic variants
that do not influence the exposures in very similar ways.

Instrument strength is usually measured using F-statistics. In the
univariable setting with only one exposure, F-statistics from a linear
regression of the exposure on genetic variants are often reported,
with larger values of the F-statistic signalling stronger instruments.
A convenient rule-of-thumb following the influential findings of \citet{Staiger1997}
is that an F-statistic below 10 is cause for concern that an analysis
may be unreliable due to weak instrument bias. In the multivariable
setting, conditional F-statistics \citep{Sanderson2016} are a useful
measure of conditional instrument strength, and again a rule-of-thumb
threshold of 10 has been suggested to indicate sufficient conditional
instrument strength \citep{Sanderson2019}. 

Of the 50 most cited articles containing the words ``multivariable
Mendelian randomization'' in the title or abstract in the Scopus
database on 27th June 2024, at least 30 make reference to F-statistics
(either univariable or conditional), at least 20 make reference to
the rule-of-thumb threshold for F-statistics being 10, and at least
19 make reference explicitly to conditional F-statistics. This brief
survey suggests that the dangers of weak instruments are well appreciated
in multivariable Mendelian randomization practice. 

However, in the reporting of empirical results, we might be facing
an example of Goodhart's law in practice: \emph{When a measure becomes
a target, it ceases to be a good measure}. If studies are selecting
instruments based on demonstrating large enough values of F-statistics,
and choosing not to report results with low F-statistics, then this
may lead to a type of publication bias that has been discussed in
epidemiology \citep{Burgess2011} and economics \citep{Zivot1998,Andrews2019}. 

Low values of conditional F-statistics are not in themselves a reason
to avoid performing a multivariable Mendelian randomization analysis.
Instead, we could implement \textsl{weak instrument-robust} strategies
\citep{Moreira2003} that provide valid confidence sets even in weak
instrument scenarios. Such approaches have been considered in univariable
Mendelian randomization analyses \citep{Wang}, but not yet in the
multivariable setting despite a clear need. In multivariable Mendelian
randomization, estimation bias is not necessarily towards a null hypothesis
of no exposure effect even in the two-sample setting, and thus can
lead to inflated type I error rates. 

In this paper, we develop two-sample summary data versions of \textsl{multivariable}
weak instrument-robust test statistics based on \citet{Anderson1949},
\citet{Kleibergen2005}, and \citet{Andrews2018}. We apply the approach
of \citet{Andrews2018}, which encourages the reporting of both a
non-robust and weak instrument-robust confidence set, along with a
so-called coverage \textsl{distortion cutoff} that measures the maximum
coverage loss of the non-robust set under weak instruments. This distortion
cutoff is perhaps a more intuitive measure of instrument strength
than the conditional F-statistic, and leaves the decision of which
confidence set should be prioritised to the reader. If the distortion
cutoff is deemed too high, then we will believe only the robust confidence
set, whereas if it is deemed acceptable then we will also consider
the non-robust confidence set. 

Our main focus is on weak instrument-robust inference, but currently
such methods are not robust to invalid instruments. Therefore, to
make inferences robust to both weak instruments and overdispersion
heterogeneity (where instruments may have direct effects on the outcome,
not via included exposures, analogously to a random-effects model;
see, for example, \citealp{Bowden2017}), we derive a novel adjustment
to \citet{Kleibergen2005} statistics that recovers high coverage
rates under overdispersion heterogeneity. 

For straightforward application of the methods we discuss, we have
developed an R package (available at \href{http://www.github.com/ash-res/mvmr-weakiv/}{github.com/ash-res/mvmr-weakiv/})
for use in multivariable Mendelian randomization studies with two-sample
summary data. We encourage the use of these methods instead of practices
such as screening on F-statistics that can do more harm than good
\citep{Angrist2024}. As an empirical illustration, we consider mean
arterial pressure and pulse pressure as two correlated exposures,
and estimate their direct effects on stroke incidence.

\section{Model and methods under strong instruments}

\subsection{Linear model with two-sample summary data}

The majority of multivariable Mendelian randomization methods work
with a summary data version of a linear instrumental variable model
with homoskedastic errors. Let $Y$ denote an outcome, $X=(X_{1},\ldots,X_{K})^{\prime}$
a $K$-vector of exposures, and $Z=(Z_{1},\ldots,Z_{J})^{\prime}$
a $J$-vector of genetic variants that are valid instruments for the
exposures, where $J\geq K$. Our goal is to make accurate inference
on the $K$-vector of exposure effects $\theta_{0}=(\theta_{1},\ldots\theta_{K})^{\prime}$
on the outcome, where, for example, $\theta_{1}$ represents the effect
of varying $X_{1}$ while holding the other exposures $X_{2},\ldots,X_{K}$
constant; i.e.\@ $\theta_{0}$ is a vector of the controlled direct
effect of each exposure. 

If the variant effects on the exposure, and the exposure effects on
the outcome, can be described by linear models, then we have the summary
data model 
\begin{equation}
\Gamma=\gamma\theta_{0}
\end{equation}
where $\Gamma=var(Z)^{-1}cov(Z,Y)$ is the $J$-vector of coefficients
from the population regression of $Y$ on $Z$, and $\gamma$ is the
$J\times K$ matrix where its $k$-th column $\gamma_{k}=var(Z)^{-1}cov(Z,X_{k})$
is the $J$-vector of coefficients from a population regression of
$X_{k}$ on $Z$. 

To make inferences on $\theta_{0}$, we need to construct estimates
of $\Gamma$ and $\gamma$ using only summary data from univariable
$Y$ on $Z_{j}$, and $X_{k}$ on $Z_{j}$ linear regressions typically
reported in genome-wide association studies. We work with the popular
two-sample setting where summary data from $X_{k}$ on $Z_{j}$ regressions
are computed from an $n_{X}$-sized sample that is non-overlapping
with a $n_{Y}$-sized sample that is used to compute summary data
from $Y$ on $Z_{j}$ linear regressions. We also assume that $n_{X}\big/n_{Y}\to c$
for some constant $c>0$ as $(n_{X},n_{Y})\to\infty$, so that the
sampling uncertainty from both summary datasets is not ignorable. 

Additionally to genetic association data, we also require good estimates
of the $J\times J$ genetic variant correlation matrix, and the $K\times K$
exposure correlation matrix, for the general case where we want to
perform an analysis with correlated variants and correlated exposures.
By similar arguments used in \citet{Wang} and \citet{Patel2024},
we can construct \textsl{multivariable} estimates $(\widehat{\Gamma},\widehat{\Sigma}_{\Gamma})$
and $(vec(\widehat{\gamma}),\widehat{\Sigma}_{\gamma})$ that can
be modelled in large samples as 
\begin{equation}
\sqrt{n_{X}}\begin{bmatrix}\widehat{\Gamma}-\gamma\theta_{0}\\
vec(\widehat{\gamma})-vec(\gamma)
\end{bmatrix}\sim N\left(\begin{bmatrix}0_{J\times1}\\
0_{JK\times1}
\end{bmatrix},\widehat{\Sigma}=\begin{bmatrix}\widehat{\Sigma}_{\Gamma} & 0_{J\times JK}\\
0_{JK\times J} & \widehat{\Sigma}_{\gamma}
\end{bmatrix}\right)
\end{equation}
where for any $J\times K$ matrix $A$, $vec(A)$ denotes the $JK$-vector
formed by stacking the columns of $A$ (see Supplementary Material
for further details). The aim is now to use the normality approximation
in Equation (2) to construct good confidence sets for $\theta_{0}$. 

\subsection{GMM estimation with strong instruments}

Under strong instruments, various multivariable Mendelian randomization
methods have been designed to offer some robustness to invalid instruments.
Since our focus in this paper is not on invalid instruments, we do
not review those here. A widely adopted estimator is based on the
two-sample multivariable inverse variance weighted method, which under
some regularity conditions is asymptotically equivalent to the two-stage
least squares estimator with one-sample data \citep{Burgess2015a}. 

A natural alternative are generalized method of moments (GMM; \citealp{Hansen1996})-based
methods that have been shown to offer \textsl{some} robustness to
weak instruments \citep{Davies2015}, as well as possibly being more
efficient (i.e.\@ having a lower variance in estimation). To describe
this estimation strategy, we can use Equation (2) to construct a $J$-vector
of moment functions, $\widehat{g}(\theta)=\widehat{\Gamma}-\widehat{\gamma}\theta$.
Setting $\widehat{g}(\theta)=0_{J\times1}$ gives $J$ estimating
equations for the $K$ unknown parameters $\theta$. Therefore, when
$J>K$ we typically cannot find a unique estimator $\widetilde{\theta}$
that sets all the estimating equations to zero, $\widehat{g}(\widetilde{\theta})=0_{J\times1}$.
Hence, GMM methods look to minimise a criterion that measures how
close $\widehat{g}(\theta)$ is to $0_{J\times1}$ over different
values of $\theta$. 

An estimate of the large sample variance of the moment functions $\widehat{g}(\theta)$
is $\widehat{\Omega}(\theta)=\widehat{\Sigma}_{\Gamma}+\varphi(\theta)\widehat{\Sigma}_{\gamma}\varphi(\theta)^{\prime}$,
where $\varphi(\theta)=I_{J}\otimes\theta_{0}^{\prime}$, and $\otimes$
denotes the Kronecker product. The GMM point estimator of $\theta_{0}$
is then defined as $\widehat{\theta}=\arg\min_{\theta}\widehat{Q}(\theta)$,
where $\widehat{Q}(\theta)=\widehat{g}(\theta)^{\prime}\widehat{\Omega}(\theta)^{-1}\widehat{g}(\theta)$.
When instruments are sufficiently strong, the GMM estimator $\widehat{\theta}$
will be consistent for $\theta_{0}$ and asymptotically normal. An
estimator of the large sample variance of $\widehat{\theta}$ is $\widehat{\Sigma}_{\widehat{\theta}}=(\widehat{\gamma}^{\prime}\widehat{\Omega}(\widehat{\theta})^{-1}\widehat{\gamma})^{-1}$,
so that the square root of the $k$-th diagonal element of $\widehat{\Sigma}_{\widehat{\theta}}$
divided by $\sqrt{n_{X}}$ can be used as the standard error of the
$k$-th exposure effect estimate $\widehat{\theta}_{k}$. 

\subsection{Non-robust confidence sets using a Wald statistic}

Using the GMM estimator $\widehat{\theta}$, and its large sample
variance estimator $\widehat{\Sigma}_{\widehat{\theta}}$ described
above, one way to construct confidence sets is through Wald test statistics,
${\cal W}(\theta)=n_{X}(\widehat{\theta}-\theta)^{\prime}\widehat{\Sigma}_{\widehat{\theta}}^{-1}(\widehat{\theta}-\theta)$.
If we were interested in testing the null hypothesis $H_{0}:\theta=\theta_{0}$
that the true exposure effects are equal to a certain value $\theta_{0}$,
then under $H_{0}$ and in large samples, the Wald statistic ${\cal W}(\theta_{0})$
should behave like a $\chi_{K}^{2}$ random variable, a chi-squared
random variable with $K$ degrees of freedom. If the true value is
different from $\theta_{0}$, then we would expect $\widehat{\theta}-\theta_{0}$
to drift away from $0_{K\times1}$ so that ${\cal W}(\theta_{0})$
becomes large. 

Using this idea, confidence sets for the true exposure effects can
be constructed by \textsl{inverting} the Wald test, which simply means
collecting the values of $\theta_{0}$ that are not rejected by the
Wald test. If $\chi_{K,1-\alpha}^{2}$ is the $1-\alpha$ quantile
of the $\chi_{K}^{2}$ distribution, then a $(1-\alpha)\times100\%$
confidence set $CS_{W}$ collects the values $\theta$ such that ${\cal W}(\theta)$
is less than or equal to $\chi_{K,1-\alpha}^{2}$. This confidence
set will perform well in terms of coverage when instruments are strong,
but this is a \textsl{non-robust} confidence set in that it is not
expected to achieve nominal coverage probability $1-\alpha$ when
instruments are weak. 

\subsection{Measuring instrument strength using conditional F-statistics}

Whether we use the inverse variance weighted method, the GMM method
outlined in Section 2.2, invert the Wald statistic as in Section 2.3,
or others that compute confidence intervals using a normality assumption,
making reliable inferences will require \textsl{conditionally} strong
instruments. In contrast with univariable analyses with just one exposure,
it is not enough that we have strong genetic associations with each
exposure, but we also require that these genetic associations are
unique. Specifically, we require that genetic associations with each
exposure are linearly independent.

We can measure conditional instrument strength using conditional F-statistics
\citep{Sanderson2016}. For each exposure $k$, the conditional F-statistic
measures the extent to which the genetically-predicted exposure effects
$\gamma_{k}$ can be expressed as a linear combination of other genetically-predicted
exposure effects $\gamma_{m},(m=1,\ldots,K,m\not=k)$ in the model.
Thus, we calculate one conditional F-statistic for each exposure in
the model. 

To construct conditional F-statistics using summary data, we require
some additional notation. Let $\widehat{\Sigma}_{\gamma,k}$ denote
the matrix $\widehat{\Sigma}_{\gamma}$ that is rearranged such that
the $k$-th column is moved left to be the first column, and the $k$-th
row is moved up to be the first row. Let $\widehat{\gamma}_{-k}$
denote the $J\times(K-1)$ matrix equal to $\widehat{\gamma}$ without
the $k$-th column, let $\gamma_{-k}$ denote the $J\times(K-1)$
matrix equal to $\gamma$ without the $k$-th column, and let $h(\delta)=[I_{J}\,\,\,-I_{J}\otimes\delta^{\prime}]$
denote the $J\times JK$ matrix where $\delta$ is a $(K-1)$-vector
of unknown parameters. 

Then, for any exposure $k$, we consider conditional F-statistics
given by 
\[
F_{k\vert-k}=\frac{n_{X}}{J-K+1}\min_{\delta}\{(\widehat{\gamma}_{k}-\widehat{\gamma}_{-k}\delta)^{\prime}[h(\delta)\widehat{\Sigma}_{\gamma,k}h(\delta)^{\prime}]^{-1}(\widehat{\gamma}_{k}-\widehat{\gamma}_{-k}\delta)\}.
\]
If we can express the $k$-th genetically-predicted exposure effect
as a linear combination of other genetically-predicted exposure effects,
so that $\gamma_{k}=\gamma_{-k}\delta$, then the conditional F-statistic
$F_{k\vert-k}$ multiplied by $J-K+1$ should behave like a $\chi_{J-K+1}^{2}$
random variable in large sample sizes $n_{X}$. More generally, a
large value of a conditional F-statistic suggests we have conditionally
strong instruments for that exposure.

\section{Methods for weak instrument-robust inference}

\subsection{Modelling weak instrument scenarios}

When instruments are weak, inference on $\theta_{0}$ using the strategies
discussed in Section 2 will not be valid. This is because under weak
instruments, the large sample distributions of conventional test statistics,
such as the Wald statistic, become more complicated. In particular,
these statistics follow nonstandard distributions that depend on instrument
strength, and hence simply comparing these test statistics against
critical values from the standard normal or chi-squared distributions
will not yield valid inference. 

To model the impact of weak instruments, methods have used the technique
of describing instrument effects on the exposures $\gamma$ as \textsl{local-to-zero},
decreasing at some rate with the sample size $n_{X}$. This modelling
technique is known as weak instrument asymptotics. Weak instrument-robust
methods are based on weak instrument asymptotics that are typically
similar to those considered in \citet{Staiger1997}, describing a
severe weak instruments problem for a small number of instruments. 

These rate restrictions are different to \textsl{many weak} instrument
asymptotics that model the scenario that many instruments have small
effects on the exposures, allowing point estimation of exposure effects
but requiring that standard errors are adjusted to account for the
uncertainty due to many weak instruments \citep{Newey2009a,Zhao2018}.
In contrast, weak instrument-robust methods do not provide point estimates,
but only confidence sets. However, given that inference, rather than
point estimation, is the primary goal in Mendelian randomization studies,
this should not be considered a disadvantage. 

If we consider weak instrument asymptotics of the form $\gamma=\mu\big/\sqrt{n_{X}}$,
where $\mu$ is a fixed full rank $J\times K$ matrix, then large
sample arguments justify the normality assumption 
\begin{equation}
\sqrt{n_{X}}\begin{bmatrix}\widehat{g}(\theta_{0})\\
vec(\widehat{\gamma})
\end{bmatrix}\sim N\left(\begin{bmatrix}0_{J\times1}\\
vec(\mu)
\end{bmatrix},\begin{bmatrix}\widehat{\Omega}(\theta_{0}) & \widehat{\Delta}(\theta_{0})\\
\widehat{\Delta}(\theta_{0})^{\prime} & \widehat{\Sigma}_{\gamma}
\end{bmatrix}\right),\,\,\text{where}\,\,\widehat{\Delta}(\theta_{0})=-\varphi(\theta_{0})\widehat{\Sigma}_{\gamma}.
\end{equation}
Note that the joint distribution of the moment functions $\widehat{g}(\theta_{0})$
and the estimated genetically-predicted exposure effects $\widehat{\gamma}$
depends on a $J\times K$ matrix of unknown nuisance parameters $\mu$.
As a result, the large sample distributions of conventional test statistics
that we use to construct confidence sets for $\theta_{0}$ will depend
on these nuisance parameters under weak instruments. The weak instrument-robust
methods we describe below seek to construct test statistics that in
some way avoid dependence on $\mu$. 

\subsection{Anderson and Rubin (1949) statistics}

Perhaps the simplest approach for weak instrument-robust inference
works with the \citet{Anderson1949} statistic. If we reconsider the
GMM criterion $\widehat{Q}(\theta)$ defined in Section 2.2, instead
of trying to find a point estimator $\widehat{\theta}$ that minimises
this criterion, we can note that for the true value of exposure effects
$\theta_{0}$, the statistic ${\cal S}(\theta_{0})=n_{X}\widehat{Q}(\theta_{0})$
behaves like a $\chi_{J}^{2}$ random variable in large samples. 

Importantly, this limiting distribution holds regardless of the value
of instrument strength $\mu$, and therefore we can invert a test
based on the statistic ${\cal S}(\theta)=n_{X}\widehat{Q}(\theta)$
to construct robust confidence sets analogously to $CS_{W}$ in Section
2.3. A $(1-\alpha)\times100\%$ Anderson-Rubin confidence set $CS_{AR}$
for $\theta_{0}$ collects the values of $\theta$ such that ${\cal S}(\theta)$
is less than or equal to $\chi_{J,1-\alpha}^{2}$. Since the degrees
of freedom is increasing with the number of instruments $J$, when
$J$ far exceeds the number of exposures $K$, $CS_{AR}$ may result
in conservative inference (Figure S2, Supplementary Material). 

If we look closely at the Anderson-Rubin statistic ${\cal S}(\theta)=n_{X}\widehat{Q}(\theta)$,
we can see that this is related to ``Q-statistic'' heterogeneity
tests in Mendelian randomization \citep{Bowden2018}, and overidentification
tests \citep{Hansen1982} more generally, that are often used to assess
the coherency of evidence across multiple instruments. Thus, the Anderson-Rubin
test doubles as a type of overidentification test. If the confidence
set $CS_{AR}$ is empty, this suggests that no potential values of
$\theta_{0}$ may be compatible with the model given the observed
data, which may be due to excessive heterogeneity in genetic variant--outcome
associations. 

Although this built-in check of excessive heterogeneity appears to
be a useful feature of Anderson-Rubin confidence sets, \citet{Davidson2014}
discuss how this can lead to overly precise confidence sets when $\min_{\theta}{\cal S}(\theta)$
is less than but close to the critical value $\chi_{J,1-\alpha}^{2}$;
ths reason the confidence set could be small is not only because instruments
are strong, but potentially because relatively few potential values
of $\theta_{0}$ may be deemed acceptable according to an overidentification
test.

\subsection{Kleibergen (2005) statistics}

A potentially more powerful approach could aim to incorporate information
on instrument strength. To do this while removing the dependence on
nuisance parameters $\mu$, \citet{Moreira2003} proposes tests that
condition on a sufficient statistic for $\mu$ that is uncorrelated
with $\widehat{g}(\theta_{0})$ in large samples. To construct such
a sufficient statistic, we start by partitioning $\widehat{\Delta}(\theta_{0})=[\widehat{\Delta}_{1}(\theta_{0})\,\,\widehat{\Delta}_{2}(\theta_{0})\,\,\ldots\,\widehat{\Delta}_{K}(\theta_{0})]$
where, for example, $\widehat{\Delta}_{1}(\theta_{0})$ corresponds
to the leftmost $J\times J$ block in $\widehat{\Delta}(\theta_{0})$,
$\widehat{\Delta}_{2}(\theta_{0})$ the next $J\times J$ block, and
so on. For each exposure $k$, let $\widehat{D}_{k}(\theta_{0})=\widehat{\gamma}_{k}-\widehat{\Delta}_{k}(\theta_{0})^{\prime}\widehat{\Omega}(\theta_{0})^{-1}\widehat{g}(\theta_{0})$.
Then, a sufficient statistic for $\mu$ is given by $\widehat{D}(\theta_{0})$,
which is the $J\times K$ matrix such that its $k$-th column is $\widehat{D}_{k}(\theta_{0})$.
Moreover, $\widehat{D}_{k}(\theta_{0})$ is uncorrelated with $\widehat{g}(\theta_{0})$
in large samples, as required. 

\citet{Kleibergen2005} considers inference on $\theta_{0}$ using
the statistic ${\cal K}(\theta_{0})=n_{X}\widehat{g}(\theta_{0})^{\prime}\widehat{\Omega}(\theta_{0})^{-1}\widehat{D}(\theta_{0})$
$\times[\widehat{D}(\theta_{0})^{\prime}\widehat{\Omega}(\theta_{0})\widehat{D}(\theta_{0})]^{-1}\widehat{D}(\theta_{0})^{\prime}\widehat{\Omega}(\theta_{0})^{-1}\widehat{g}(\theta_{0})$,
and shows that this statistic behaves like a $\chi_{K}^{2}$ random
variable even under weak instrument asymptotics. As before, test inversion
gives a weak instrument-robust confidence set, so that a $(1-\alpha)\times100\%$
confidence set $CS_{K}$ for $\theta_{0}$ collects the values $\theta$
such that ${\cal K}(\theta)$ is less than or equal to $\chi_{K,1-\alpha}^{2}$. 

\subsection{Andrews (2018) statistics}

One reason why weak instrument-robust methods have not been widely
adopted in Mendelian randomization studies so far may be their conservatism,
i.e.\@ the robust confidence sets for the exposure effects in Sections
3.2 and 3.3 may be a lot larger than conventional non-robust confidence
sets even under quite strong instruments. In applied practice, investigators
may choose to report a non-robust confidence set if instruments are
demonstrably strong (for example, by considering F-statistics), and
otherwise report weak instrument-robust confidence sets. \citet{Andrews2018}
studies this strategy, and proposes the reporting of both robust and
non-robust confidence sets, along with a coverage \textsl{distortion
cutoff} statistic that gives some sense of how reliable the non-robust
confidence set is in terms of coverage. 

The non-robust confidence set $CS_{N}$ considered by \citet{Andrews2018}
also inverts a Wald test as in Section 2.3, but the Wald statistic
is slightly different to ${\cal W}(\theta)$. In particular, the $(\widehat{\theta}-\theta)$
term is replaced by $(\bar{\theta}(\theta)-\theta)$ where $\bar{\theta}(\theta)=\theta+(\widehat{\gamma}^{\prime}\widehat{\Omega}(\theta)^{-1}\widehat{\gamma})^{-1}\widehat{\gamma}^{\prime}\widehat{\Omega}(\theta)^{-1}\widehat{g}(\theta)$
behaves similarly to the GMM estimator $\widehat{\theta}$ in large
samples, provided $\theta$ lies quite close to the true value $\theta_{0}$. 

Likewise, for weak instrument-robust inference, the \citet{Kleibergen2005}-type
statistic considered by \citet{Andrews2018} is also slightly different
to ${\cal K}(\theta)$ in Section 3.3, and is defined as ${\cal K}^{\star}(\theta)=n_{X}(\theta^{\star}(\theta)-\theta)^{\prime}(\widehat{D}(\theta)^{\prime}\widehat{\Omega}(\theta)^{-1}\widehat{D}(\theta))^{-1}(\theta^{\star}(\theta)-\theta)$
where $\theta^{\star}(\theta)=\theta+(\widehat{D}(\theta)^{\prime}\widehat{\Omega}(\theta)^{-1}\widehat{D}(\theta))^{-1}\widehat{D}(\theta)^{\prime}\widehat{\Omega}(\theta)^{-1}\widehat{g}(\theta)$. 

This Kleibergen statistic ${\cal K^{\star}}(\theta_{0})$ behaves
like a $\chi_{K}^{2}$ random variable in large samples, and is asymptotically
independent of the statistic ${\cal S}(\theta_{0})-{\cal K}^{\star}(\theta_{0})$
which behaves like a $\chi_{J-K}^{2}$ random variable in large samples
\citep[Thm 2,  p.342]{Andrews2018}. This suggests a linear combination
of the Anderson-Rubin and Kleibergen statistics, ${\cal K}^{\star}(\theta_{0})+a\cdot{\cal S}(\theta_{0})$
should behave like a $(1+a)\cdot\chi_{K}^{2}+a\cdot\chi_{J-K}^{2}$
random variable in large samples for a fixed choice of $a\in[0,1]$,
where $\chi_{K}^{2}$ and $\chi_{J-K}^{2}$ are independent chi-squared
random variables. Therefore, a robust confidence set with nominal
coverage probability $1-\alpha$ may be constructed by collecting
the values $\theta$ for which ${\cal K}^{\star}(\theta)+a\cdot{\cal S}(\theta)<q(1-\alpha;a,J,K)$,
where $q(1-\alpha;a,J,K)$ is the $1-\alpha$ quantile of the distribution
of $(1+a)\cdot\chi_{K}^{2}+a\cdot\chi_{J-K}^{2}$, which can be easily
calculated by simulation. 

Given this result, a natural question is: how should $a$ be chosen
such that inference based on the linear combination statistic ${\cal K}^{\star}(\theta_{0})+a\cdot{\cal S}(\theta_{0})$
has some useful properties? \citet{Andrews2018} ties the choice of
$a=a(\gamma)$ to the maximal coverage distortion $\gamma$ of the
non-robust confidence set $CS_{N}$ under weak instruments. This approach
offers a practical decision framework for dealing with weak instruments:
investigators that are not willing to accept a coverage distortion
$\gamma$ will be interested only in weak instrument-robust confidence
set, whereas those investigators that are would also be interested
in the non-robust confidence set $CS_{N}$. 

The approach starts by defining $\gamma_{min}$ as the minimum coverage
distortion that an investigator could be willing to accept. For any
given value $\gamma\geq\gamma_{min}$, the value of $a(\gamma)$ is
solved as $q(1-\alpha-\gamma;a(\gamma),J,K)=\chi_{K,1-\alpha}^{2}$.
A $(1-\alpha)\times100\%$ robust confidence set $CS_{R}$ simply
collects the values $\theta$ for which ${\cal K}(\theta)+a(\gamma_{min})\cdot{\cal S}(\theta)$
is less than $q(1-\alpha;a(\gamma_{min}),J,K)$. This robust confidence
set $CS_{R}$ should achieve nominal coverage probability $1-\alpha$
under strong and weak instrument settings. 

To calculate maximum coverage distortions of the non-robust set $CS_{N}$,
define a confidence set $CS_{P}(\gamma)$ that collects values $\theta$
such that the statistic ${\cal K}^{\star}(\theta)+a(\gamma)\cdot{\cal S}(\theta)$
is less than $\chi_{K,1-\alpha}^{2}$. By construction of $a(\gamma)$,
this confidence set should have coverage probability of at least $1-\alpha-\gamma$
in large samples. Under some regularity conditions, \citet[Thm 3, p.344]{Andrews2018}
shows that $CS_{P}(\gamma)$ should be contained in a nominal $(1-\alpha)\times100\%$
non-robust confidence set $CS_{N}$ under a strong instruments setting.
Thus, if $CS_{P}(\gamma)$ is contained in $CS_{N}$, then $CS_{N}$
should also have coverage probability of at least $1-\alpha-\gamma$
in large samples. 

Starting from the minimum coverage distortion $\gamma_{min}$, we
can increase $\gamma$ until we find the smallest value $\widehat{\gamma}$
such that $CS_{P}(\widehat{\gamma})$ is contained in $CS_{N}$. This
value $\widehat{\gamma}$ is defined as the \textsl{distortion cutoff}
by \citet{Andrews2018}, and represents the maximum coverage distortion
of the non-robust confidence set $CS_{N}$ under weak instruments
in large samples. Moreover, $\widehat{\gamma}$ converges to $\gamma_{min}$
under strong instruments, and thus provides a useful measure of instrument
strength. 

\subsection{Kleibergen inference adjusted for overdispersion heterogeneity}

While our main focus is on making inferences that are robust to a
weak instruments problem, here we briefly consider the problem of
making inferences that are also robust to invalid instruments. A popular
class of Mendelian randomization methods are designed to be robust
to \textsl{overdispersion heterogeneity}, where instruments may have
a direct effect on the outcome, not via their effects on included
exposures, analogously to heterogeneity in a random-effects model
\citep{Bowden2017}. The weak instrument-robust methods described
above may perform unreliably under overdispersion heterogeneity, giving
overly precise confidence sets with poor coverage.

To model overdispersion heterogeneity, we augment a $J$-vector of
random-effects $\nu$ to Equation (1), so that the summary data model
to be estimated is 
\begin{equation}
\Gamma=\gamma\theta_{0}+\nu,\,\,\,\,\,\nu\sim N(0_{J\times1},I_{J}\kappa^{2}n_{Y}^{-1})
\end{equation}
for some unknown overdispersion heterogeneity parameter $\kappa^{2}\geq0$.
With overdispersion heterogeneity, the moment function $\widehat{g}(\theta_{0})$
described in Section 2.2 still has mean zero, but now a consistent
estimator for its asymptotic variance is $\widehat{\Omega}(\theta_{0},\kappa^{2})=\widehat{\Sigma}_{\Gamma}+I_{J}c\kappa^{2}+\varphi(\theta)\widehat{\Sigma}_{\gamma}\varphi(\theta)^{\prime}$,
where $n_{X}\big/n_{Y}\to c$, as $(n_{X},n_{Y})\to\infty$. 

Under the null hypothesis $H_{0}:\theta=\theta_{0}$, the quantity
$n_{X}\widehat{g}(\theta_{0})^{\prime}\widehat{\Omega}(\theta_{0},\kappa^{2})^{-1}\widehat{g}(\theta_{0})$
has mean $J$, so that $n_{X}\widehat{g}(\theta_{0})^{\prime}\widehat{\Omega}(\theta_{0},\kappa^{2})^{-1}\widehat{g}(\theta_{0})-J=0$
is an estimating equation for $\kappa^{2}$. Suppose that $\widehat{\kappa}^{2}$
solves this estimating equation. Then, by following the same steps
as in Section 3.3 but now using the plug-in asymptotic variance estimator
$\widehat{\Omega}(\theta_{0},\widehat{\kappa}^{2})$ instead of $\widehat{\Omega}(\theta_{0})$,
we can construct an adjusted Kleibergen confidence set (Kleibergen-OH)
that is robust to overdispersion heterogeneity. 

Supplementary Figures S3--S6 demonstrate that Kleibergen-OH confidence
sets achieve nominal coverage under overdispersion heterogeneity,
whereas the standard implementations of Anderson-Rubin, Kleibergen,
and Andrews confidence sets described in previous sections can have
very poor coverage. However, Kleibergen-OH confidence sets may be
conservative when using only a few weak instruments.

\subsection{Instrument selection using measures of instrument strength}

In multivariable Mendelian randomization studies, there is often the
possibility to consider many genetic variants (in some cases, in the
hundreds) as candidate instruments. It is common practice to screen
out a large majority of these candidate instruments based on measures
of F-statistics or conditional F-statistics, often in relation to
the rule-of-thumb threshold of 10. Several works have highlighted
that this practice likely causes severe size distortions using conventional
methods, and thus inaccurate reporting of empirical results \citep{Zivot1998,Burgess2011,Andrews2019,Lee2022}.
As noted in \citet[Section 4.3,  p.741]{Andrews2019}, ``\textsl{screening
on the first-stage F-statistic appears to compound, rather than reduce,
inferential problems arising from weak instruments}''. 

Nevertheless, if a large number of candidate instruments are available,
\textsl{some} data-driven instrument selection seems attractive. One
intuitive and computationally straightforward strategy is to choose
the set of instruments that maximise the minimum conditional F-statistic
across all exposures. Another strategy could repurpose the distortion
cutoff statistic of \citet{Andrews2018} discussed in Section 3.4
as a measure of instrument strength, such that instruments are selected
to minimise this distortion cutoff. In some sense, this is an abuse
of how the distortion cutoff is defined since its large sample coverage
guarantees on the non-robust confidence set may no longer hold. But
to the extent that similar data-driven instrument selection strategies
are common in practice, it is useful to consider how selection based
on conditional F-statistics and the distortion cutoff could affect
subsequent inference. We do this in Section 4.5 below. 

\section{Simulation study}

The purpose of this simulation study is to demonstrate that: (i) weak
instrument-robust methods in Section 3 achieve nominal coverage in
strong and weak instrument scenarios, whereas the non-robust approaches
discussed in Section 2 perform poorly under weak instruments (Sections
4.2, 4.3); (ii) if two-sample Mendelian randomization studies are
reported only if conditional F-statistics are at least 10 then this
likely means biased inferences among reported studies (Section 4.4).
Further, we consider the viability of two instrument selection strategies
discussed in Section 3.6 for accurate \textsl{robust} inferences (Section
4.5). 

\subsection{Design}

We generated two-sample summary data by computing univariable regression
results as described in Section 2.1, with sample sizes set to $n_{X}=n_{Y}=5000$.
For the baseline case of $J=4$ instruments and $K=2$ exposures,
the underlying linear instrumental variable model was 
\begin{eqnarray*}
Y & = & \theta_{1}X_{1}+\theta_{2}X_{2}+U\\
X_{1} & = & \gamma_{1}^{\prime}Z+V_{1}\\
X_{2} & = & \gamma_{2}^{\prime}Z+V_{2}
\end{eqnarray*}
where 
\[
\begin{bmatrix}U\\
V_{1}\\
V_{2}
\end{bmatrix}\sim N\left(\begin{bmatrix}0\\
0\\
0
\end{bmatrix},\begin{bmatrix}1 & -0.6 & 0.6\\
-0.6 & 1 & 0.3\\
0.6 & 0.3 & 1
\end{bmatrix}\right),
\]
and $Z=[Z_{1},Z_{2},Z_{3},Z_{4}]^{\prime}$ were generated from a
multivariate normal distribution with covariance matrix equal to $1$
on the diagonals, and from $aa^{\prime}$ for the off-diagonals, where
the elements of $a=[a_{1},a_{2},a_{3},a_{4}]^{\prime}$ are independent
and uniformly distributed $U[0,\sqrt{0.4}]$. This means that the
correlation between any two instruments is at most 0.4. The exposure
effects on the outcome were set at $\theta_{0}=(\theta_{1},\theta_{2})^{\prime}=(1,0)^{\prime}$,
so that exposure 1 has an effect on the outcome, but exposure 2 does
not. 

The instrument effects on exposure 1 were set at $\gamma_{1}=[1+\xi,1+\xi,1-\xi,1-\xi]^{\prime}\times0.2n_{X}^{-1/2}\mu$,
and the instrument effects on exposure 2 were set at $\gamma_{2}=[1-\xi,1-\xi,1+\xi,1+\xi]^{\prime}\times n_{X}^{-1/2}\mu$,
for different choices of \textsl{unconditional} instrument strength
$\mu>0$, and \textsl{conditional} instrument strength $\xi\in[0,1]$.
Note that when $\xi=1$, instruments $Z_{1}$ and $Z_{2}$ only affect
exposure 1, and instruments $Z_{3}$ and $Z_{4}$ only affect exposure
2, which represents the setting of high conditional instrument strength
since we have unique instruments for each exposure. Conversely, when
$\xi=0$, the instrument effects $\gamma_{1}$ and $\gamma_{2}$ are
linearly dependent, meaning we have no conditional instrument strength. 

The methods we consider for weak instrument-robust inference are the
``Anderson-Rubin'' confidence sets $CS_{AR}$ described in Section
3.2, the ``Kleibergen'' confidence sets $CS_{K}$ described in Section
3.3, and the ``Andrews'' linear combination robust confidence set
$CS_{R}$ described in Section 3.4, where the minimum acceptable coverage
distortion was set at $\gamma_{min}=0.01$. To compute confidence
sets by test inversion, we considered a grid of of potential values
formed from taking all combinations of $\theta_{1}\in[-2,2]$ and
$\theta_{2}\in[-2,2]$ in $0.04$ intervals. 

The methods we consider for non-robust inference are the multivariable
inverse variance-weighted ``IVW'', and ``GMM'', point estimators
of $(\theta_{1},\theta_{2})^{\prime}$ and related confidence intervals
discussed in Section 2.2, and the non-robust Wald-based confidence
sets $CS_{N}$ discussed in Section 3.4. 

\subsection{Bias in point estimates under weak instruments}
\begin{center}
\includegraphics[width=16.5cm]{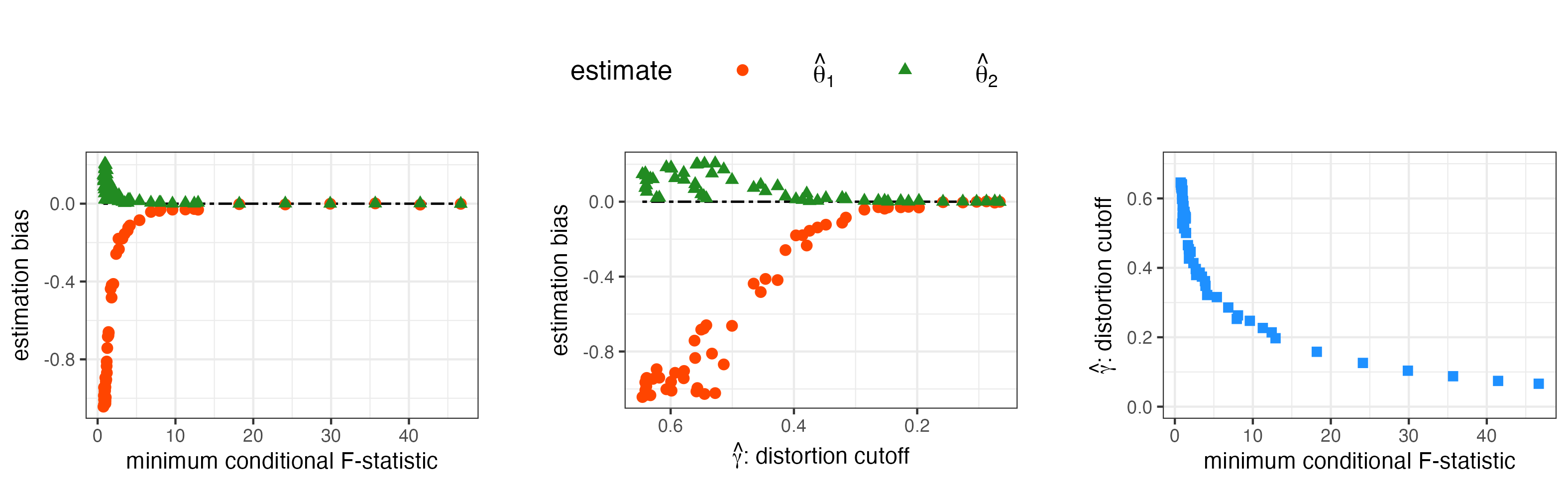}~\\
{\small{}Figure 1.} {\footnotesize{}Bias in GMM point estimates plotted
against the minimal conditional F-statistic taken over the two exposures,
and the distortion cutoff $\widehat{\gamma}$. The true exposure effect
values were $\theta_{0}=(\theta_{1},\theta_{2})=(1,0)$. }{\footnotesize\par}
\par\end{center}

Figure 1 shows how the bias in GMM point estimates of $\theta_{0}=(\theta_{1},\theta_{2})=(1,0)$
varies with two measures of instrument strength: the minimum conditional
F-statistic taken over the two exposures, and the distortion cutoff.
The right panel of Figure 1 highlights the inverse relationship between
conditional F-statistics and the distortion cutoff. The biases of
GMM point estimates are decreasing with higher values of the minimum
conditional F-statistic, and with lower values of the distortion cutoff. 

In this simulation design, the OLS point estimate (from a linear regression
of $Y$ on $X$) of $\theta_{1}$ is downward biased, whereas the
estimate of $\theta_{2}$ is upward biased. Figure 1 therefore shows
that under weak instruments, the GMM point estimates are biased in
the same direction as OLS estimates in this design. Unfortunately,
we see here that weak instrument bias is \textsl{not} necessarily
towards zero in two-sample multivariable Mendelian randomization,
which could potentially cause inflated type I error rates, i.e.\@
the greater possibility of erroneously concluding there is a non-zero
exposure effect on the outcome when there is no effect. The IVW estimates
are not plotted because there was very little difference in the performance
of IVW and GMM estimates under a weak or strong instruments setting. 

\subsection{Reliable inference using weak instrument-robust methods}
\begin{center}
\includegraphics[width=16.5cm]{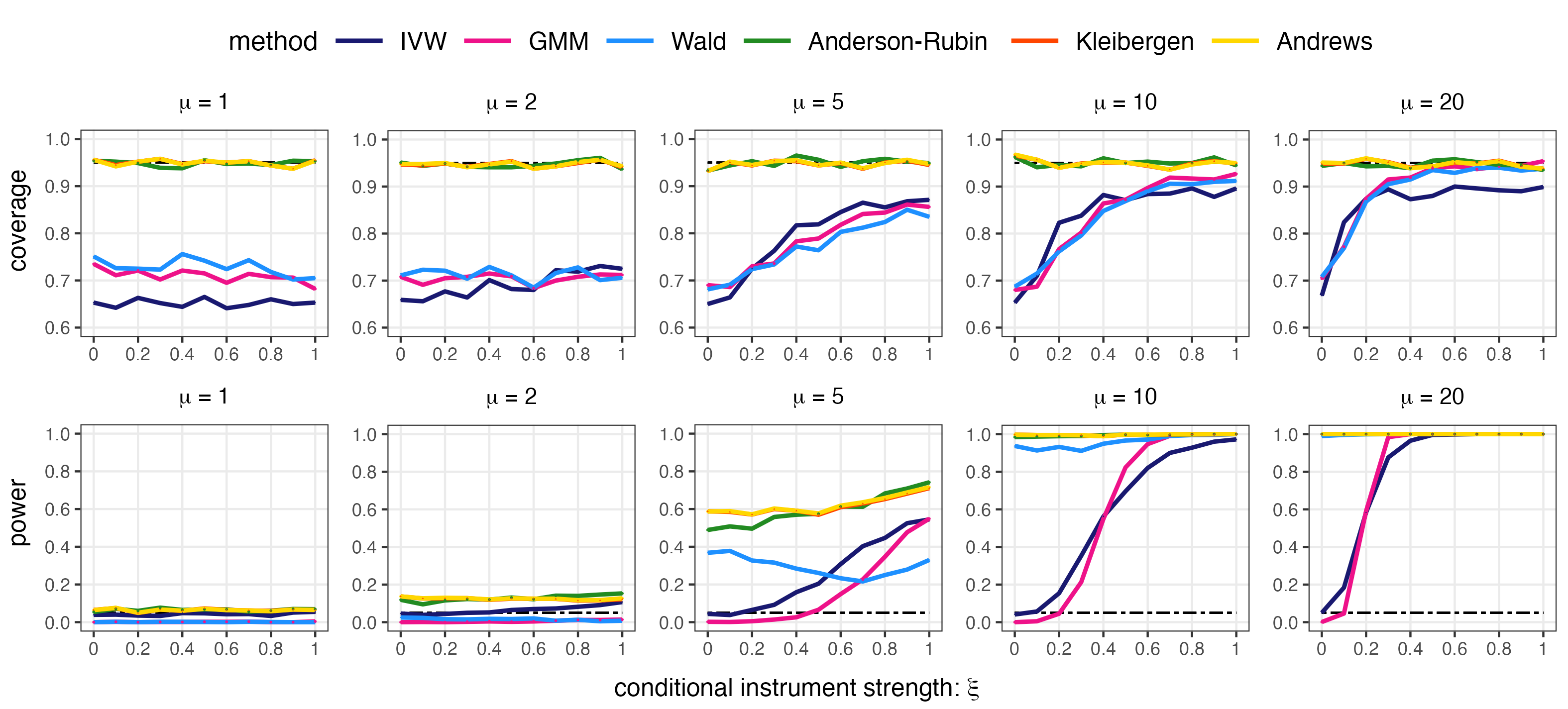}~\\
{\small{}Figure 2. }{\footnotesize{}Top row plots the coverage probability
that a 95\% confidence set contains the true value $\theta_{0}=(1,0)$,
whereas the bottom row plots the power that a 95\% confidence set
does }\textsl{\footnotesize{}not}{\footnotesize{} contain the null
value $(0,0)$. For GMM and IVW, the coverage probability that a 95\%
confidence interval for exposure 1 contains $\theta_{1}=1$, and the
power that it does not contain $0$, is plotted. }{\footnotesize\par}
\par\end{center}

Figure 2 plots how the coverage and power of 95\% confidence sets
(and 95\% confidence intervals for exposure 1 in the case of IVW and
GMM) varies with unconditional instrument strength $\mu$ and conditional
instrument strength $\xi$. All the weak instrument-robust methods
achieve coverage probability very close to the nominal 95\% level,
regardless of whether instruments are weak or strong. There is nothing
really separating the weak instrument-robust methods in terms of power
performance either, although we may expect the Anderson-Rubin confidence
sets to be more conservative if we had a larger number of instruments
(Figure S2, Supplementary Material). 

In contrast, the non-robust methods (IVW, GMM, and Wald) all have
poor coverage performance under weak instruments, and especially when
instruments are conditionally weak ($\xi$ is close to 0). The non-robust
confidence intervals for $\theta_{1}$ appear to be centered around
a much lower value than the truth $\theta_{1}=1$, so that in this
design, weak instruments has harmed the power of non-robust methods
to reject a hypothesis that $\theta_{1}$ is non-zero. Although we
do not usually think of weak instrument-robust methods as being more
powerful than non-robust methods, these results illustrate that they
can be if weak instrument bias is toward the null of no exposure effects. 

\subsection{Coverage distortions from screening on conditional F-statistics}
\begin{center}
\includegraphics[width=16.5cm]{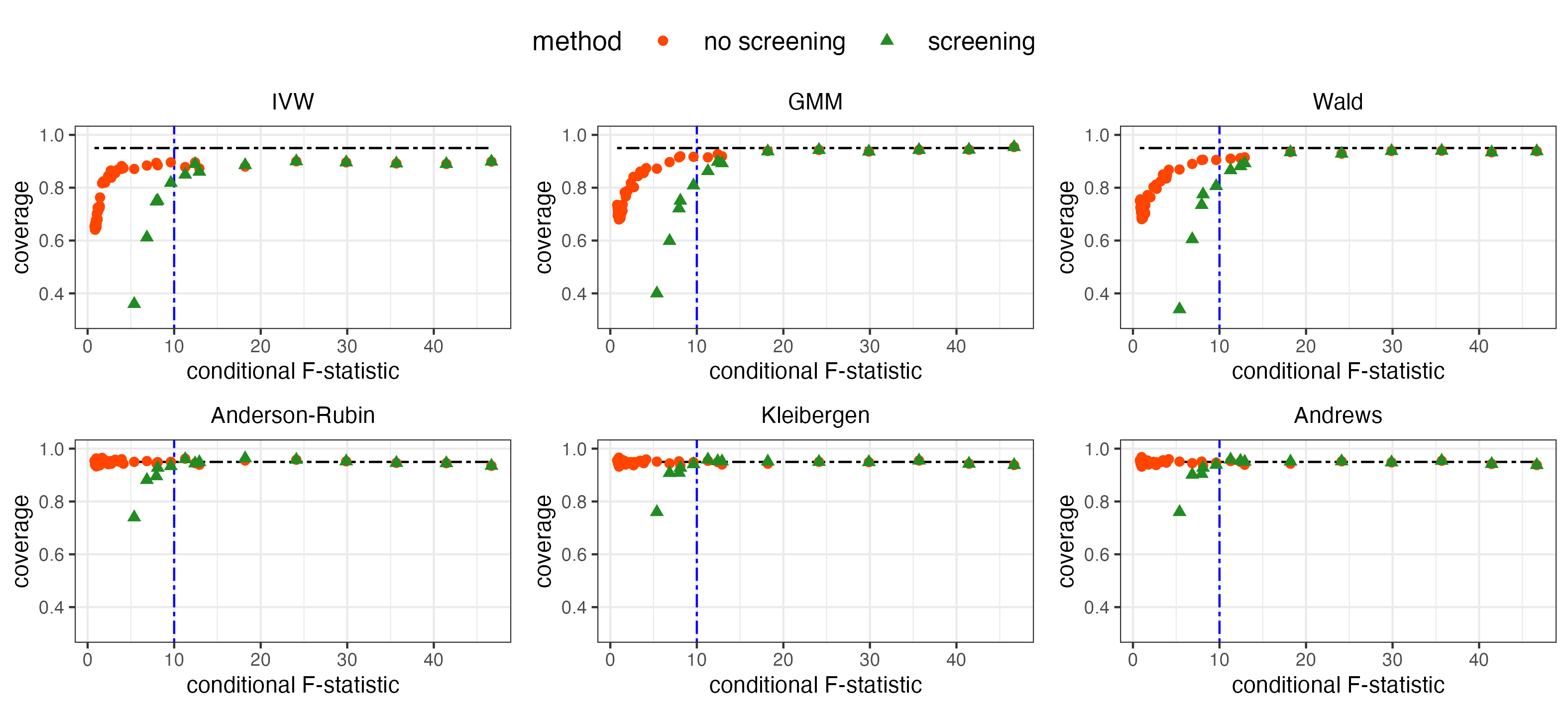}~\\
{\small{}Figure 3.} {\footnotesize{}Coverage of 95\% confidence sets
if all results are reported (no screening) versus if only those results
with the minimum F-statistic greater than 10 are reported (screening).
Each circle represents an average taken over 1000 simulations under
no screening. The x-axis plots an average minimum conditional F-statistic
taken over 1000 simulations under no screening. }{\footnotesize\par}
\par\end{center}

Now we investigate the potential impact of not reporting results if
the minimum conditional F-statistic across all exposures is under
10. Figure 3 plots the coverage of 95\% confidence intervals for $\theta_{1}$
in the case of IVW and GMM, and 95\% confidence sets for $\theta_{0}=(\theta_{1},\theta_{2})$
for the other methods, varying with the minimum conditional F-statistic
over the two exposures. As in Figure 2, the non-robust methods in
the top row have poor coverage under weak instruments, but screening
on the minimum conditional F-statistic being at least 10 can make
matters a lot worse. For example, the top-left plot of Figure 3 shows
that when the average minimum conditional F-statistic was around 6,
the coverage of the IVW-based confidence interval was nearly 90\%
without screening, whereas among screened results the coverage drops
to below 40\%. 

Similar behaviour can be seen for robust confidence sets too. Without
screening, weak instrument-robust methods control coverage even under
low conditional F-statistics, but if results where a conditional F-statistic
is less than 10 are not reported then close-to-nominal coverage rates
are no longer guaranteed. We focused on the conditional F-statistic
threshold of 10 because that is the rule-of-thumb that Mendelian randomization
applications seem to focus on, but we would expect similar problems
for other threshold choices.

\subsection{Inference after selection that maximises instrument strength}

The results in Sections 4.3 show that weak instrument-robust methods
provide reliable performance under both strong and weak instrument
scenarios. However, the choice of instruments still has a big impact
on the performance of these methods, and the use of multiple instruments
can improve precision if they are sufficiently strong. 

Here we demonstrate the utility of two instrument selection strategies:
one that maximises the minimum conditional F-statistic across all
exposures, and one that minimises the distortion cutoff. As the right
panel of Figure 1 suggests, both of these strategies are maximising
a measure of instrument strength. We do not provide an in depth discussion
of how these strategies could be implemented if choosing from many
subsets of candidate instruments; clearly this would be a computationally
difficult problem: for example, if we had $J=6$ instruments and $K=2$
exposures, there would be 57 different instrument combinations we
could take. 

Instead, we consider a simple choice between two sets of instruments:
(i) the ``full'' set of instruments, and (ii) a set of a smaller
number of ``core'' instruments. The core set of instruments are
the 4 instruments defined as before, but with the value of unconditional
instrument strength $\mu$ set to 5. The full set of instruments contain
4 instruments in addition to the core set; these additional instruments
are constructed in the same way as the core instruments, so that the
correlation between any 2 instruments in the full set of 8 instruments
is at most 0.4. However, the true instrument effects of these additional
instruments on the exposure are a multiple $\tau$ of the core instruments,
so that $(\gamma_{5},\gamma_{6},\gamma_{7},\gamma_{8})^{\prime}=(\gamma_{1},\gamma_{2},\gamma_{3},\gamma_{4})^{\prime}\cdot\tau$
for some constant $\tau>0$. 

This simple choice between the two instrument sets represents a familiar
scenario in practice: putting aside any concerns of instrument validity,
the choice of whether we want to include additional instruments depends
on how strong they are. If $\vert\tau\vert$ is close to 0, then it
is likely that using only the core set of instruments will lead to
smaller confidence sets, whereas if $\vert\tau\vert$ is much larger
than 1, then the full set of instruments may improve precision. 

The top row of Figure 4 plots the coverage probability of robust confidence
sets. Always using only the core instruments, and always using the
full instruments, would be expected to control coverage probability
at the nominal level. Interestingly, the post-selection confidence
sets based on F-statistics and the coverage distortion also appear
to control coverage quite close to the nominal level, even though
we do not have any guarantees on this in general. 

The middle row of Figure 4 plots the areas of the confidence sets,
defined as the number of grid points of potential values of $\theta_{0}$
contained in the confidence set. When $\tau=0$ so that the additional
instruments are irrelevant, using only the core instruments offers
more precise inference than using the full set of instruments. In
this case, the instrument selection strategies, especially the one
based on the conditional F-statistic, tends to select only the core
instruments. Then, as the additional instruments get stronger with
$\mu$, the full set of instruments are more quickly detected as being
optimal using the conditional F-statistic approach than using the
distortion cutoff approach. 
\begin{center}
\includegraphics[width=16.5cm]{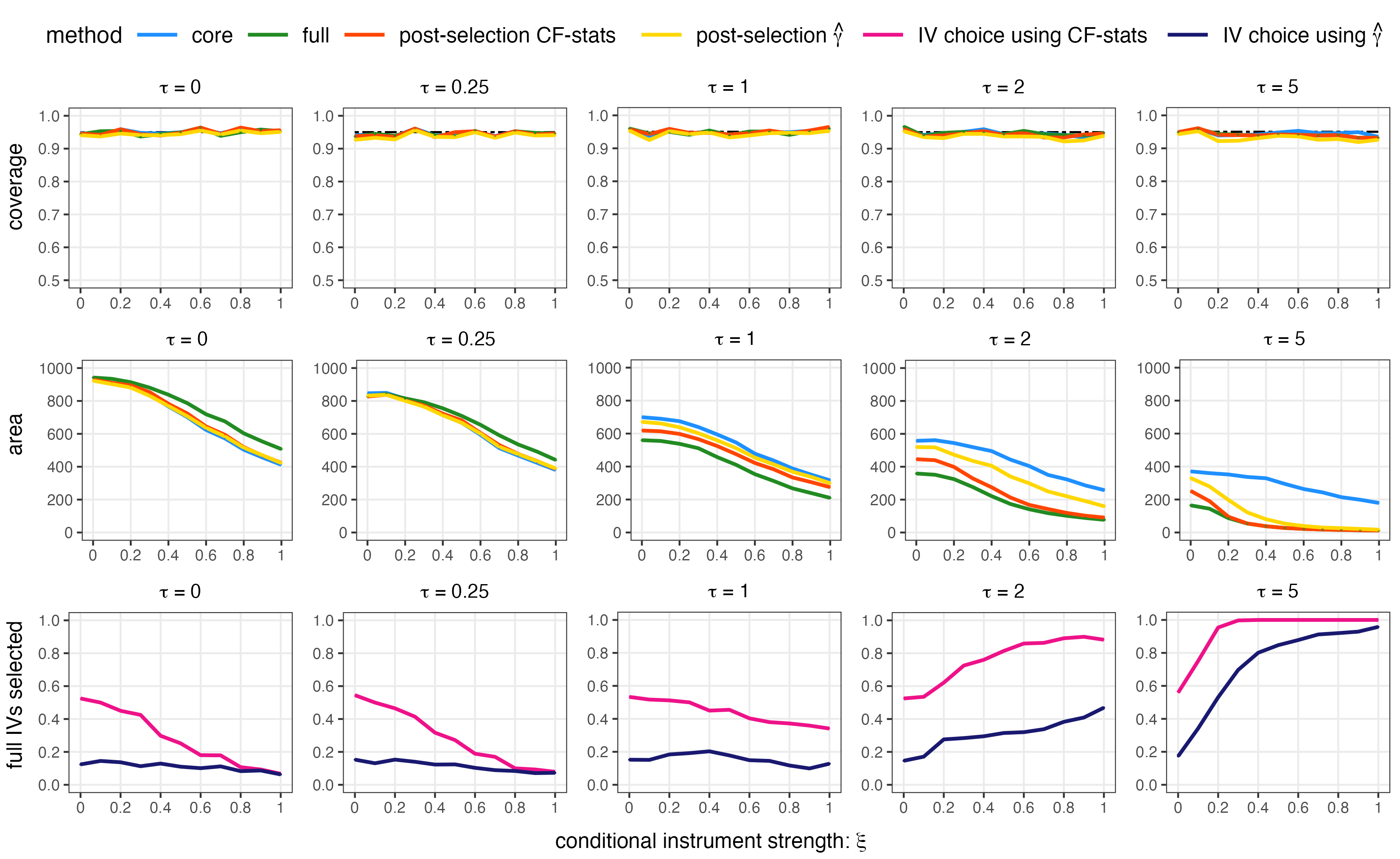}~\\
{\small{}Figure 4. }{\footnotesize{}The coverage (top row) and area
(middle row) of robust-Andrews 95\% confidence sets using: only the
core instruments (core), full set of instruments (full), instruments
that maximise the minimum conditional F-statistic (post-selection
CF stat), and instruments to minimise the distortion cutoff (post-selection
$\widehat{\gamma}$). The last row plots the rate at which the full
set of instruments were selected using the two instrument selection
strategies.}{\footnotesize\par}
\par\end{center}

Choosing only the core instruments is suboptimal if additional instruments
are strong, while choosing the full set of instruments is suboptimal
if the additional instruments are weak. Therefore, data adaptive selection
based on a measure of instrument strength could help to improve the
power of an analysis. Our results provide some support that instrument
selection strategies based on maximising the minimum conditional F-statistic,
and minimising the distortion cutoff, may be viable options when used
in conjuction with robust confidence sets. However, the extent to
which selection uncertainty affects the coverage of robust confidence
sets warrants further investigation.

\section{Empirical example: Blood pressure and stroke incidence}

A potentially challenging Mendelian randomization study is to estimate
the direct effects of mean arterial pressure (MAP) and pulse pressure
(PP) on stroke risk. Here, by the direct effect of MAP, we mean the
effect of MAP on stroke risk, holding constant the effect of PP on
stroke risk. The direct effect of PP is defined analogously. Fitting
a multivariable model in terms of MAP and PP is more interpretable
than systolic and diastolic blood pressure; MAP measures average blood
pressure, and largely reflects vascular resistance, whereas PP measures
the differential in blood pressure across the cardiac cycle, and reflects
cardiac output.

MAP is defined as ($1\big/3$ $\times$ systolic blood pressure) plus
($2\big/3$ $\times$ diastolic blood pressure), while PP is defined
as systolic blood pressure minus diastolic blood pressure. PP and
MAP are highly related, with a correlation of 0.56 in UK Biobank participants.
Therefore, it may be quite difficult to find conditionally strong
genetic predictors of both PP and MAP, especially in smaller samples. 

We consider a two-sample multivariable Mendelian randomization study,
considering MAP and PP as two exposures, and stroke incidence as the
outcome. Genetic associations with MAP and PP were computed using
a sample of $n_{X}=367,283$ participants of mostly European ancestry
in UK Biobank \citep{Sudlow2015}, and genetic associations with stroke
incidence were taken from a GWAS from the GIGASTROKE consortium \citep{Mishra2022}
involving $n_{Y}=727,571$ participants of European ancestry. We considered
the top 20 uncorrelated variants most associated with MAP, and with
PP, as instruments, so that there were at most 40 variants used as
instruments.

The 95\% non-robust (Wald) and robust (Andrews) confidence sets were
very similar using the full exposure sample $n_{X}=367,283$. The
MAP effects $[0.024,0.051]$ and the PP effects $[0.012,0.030]$ (change
in log odds ratio per unit increase in mmHg) were estimated to be
positive in the non-robust confidence set, and the results were more
precise in the robust set, $[0.031,0.049]$ for the MAP effects, and
$[0.015,0.027]$ for the PP effects. Conditional F-statistics corresponding
to MAP and PP were $73.48$ and $113.10$, and the distortion cutoff
converged to the minimum level $\widehat{\gamma}=\gamma_{min}=0.05$. 

The 95\% Anderson-Rubin confidence set was empty. However, the 95\%
Kleibergen-OH confidence set that accounts for overdispersion heterogeneity
supported the positive results of the Wald and Andrews confidence
sets, with MAP effects $[0.008,0.068]$ and PP effects $[-0.000,0.040]$.
Overall, we conclude that higher genetically-predicted MAP levels
and higher genetically-predicted PP levels are each positively associated
with higher stroke risk. 

Now we repeat this analysis but with summary data on genetic variant--exposure
associations computed using a smaller random sample from the full
sample of $n_{X}=367,283$ individuals. Specifically, we computed
confidence sets based on sample sizes $n_{X}\in\{2000,10000,20000,50000\}$.
Although this weak instrument problem is somewhat contrived, the full
sample results under strong instruments serve as a baseline for the
results we should expect as the sample size $n_{X}$ increases. Moreover,
the performance under smaller sample sizes is a relevant concern when
performing Mendelian randomization analyses in specific subgroups.

Figure 5 plots a heatmap of 100 confidence sets based on random samples
of size $n_{X}=2000$ (leftmost column) to $n_{X}=50000$ (rightmost
column), with darker contours containing values of exposure effects
most represented in the 100 confidence sets. When $n_{X}=1000$, the
average minimum conditional F-statistic over the two exposures was
$2.55$, and the average distortion cutoff was $\widehat{\gamma}=0.21$.
The non-robust confidence sets were mostly centered around the null
of no effect from either exposure, whereas the robust confidence sets
were shaped very differently. 
\begin{center}
\includegraphics[width=16.5cm]{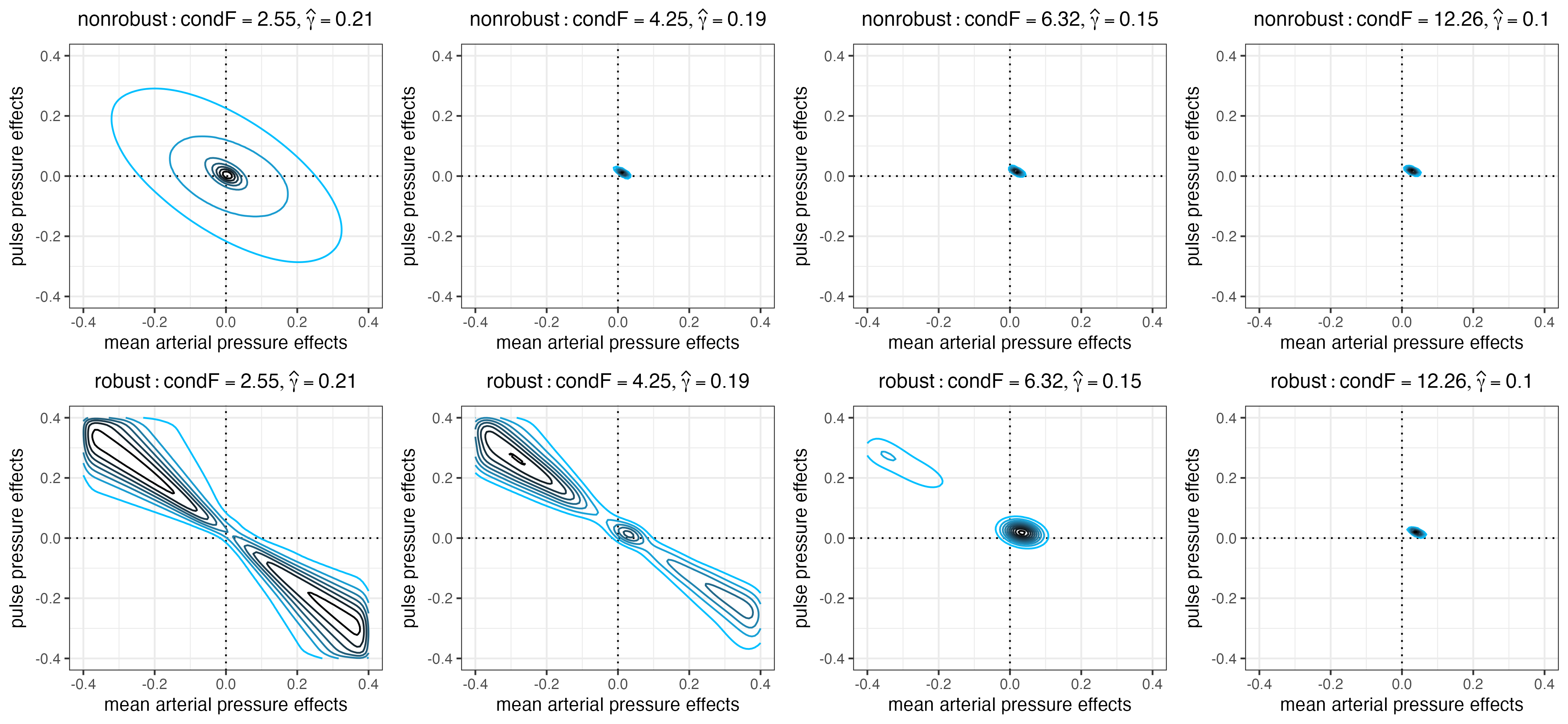}~\\
{\small{}Figure 5. }{\footnotesize{}Heatmaps of 100 non-robust (Wald;
top row) and robust (Andrews; bottom row) confidence sets of nominal
coverage 95\% for the effects of mean arterial pressure and pulse
pressure on stroke risk. The darker contours contain values of exposure
effects most represented in the 100 confidence sets. The first column
corresponds to exposure associations computed with random samples
of $n_{X}=2000$, the second column $n_{X}=10000$, the third column
$n_{X}=20000$, and the final column $n_{X}=50000$. $\widehat{\gamma}$
is the average distortion cutoff over 100 confidence sets, and condF
is the average minimum conditional F-statistic across the two exposures. }{\footnotesize\par}
\par\end{center}

The third column of Figure 5 corresponds to $n_{X}=20000$, and here
the non-robust sets are more closely centered around the region $[0.024,0.051]\times[0.012,0.030]$
estimated under the full sample. In contrast, the robust sets are
much larger (note the difference in scales). In this case, if we are
willing to accept a coverage distortion $\widehat{\gamma}=0.15$,
i.e.\@ a potential type I error rate of 20\%, then we have reason
to value the evidence from the non-robust sets, even though the minimum
conditional F-statistic was below 10 in \textsl{all} of the 100 random
samples of size $n_{X}=20000$. 

\section{Conclusion}

Multivariable Mendelian randomization applications are growing in
exciting directions; enabling studies of drug targets, time-varying
effects, mediation, disease progression, and tackling issues such
as competing risks and collider bias. Weak instrument issues are at
the heart of many of these applications, and are likely to remain
so. For example, it would seem challenging to instrument time-varying
exposures with instruments that cannot vary over time. 

Many multivariable Mendelian randomization studies are reporting conditional
F-statistics, mainly as justification that an analysis does not suffer
from weak instrument bias. While transparency on instrument strength
is useful, achieving a particular level of the conditional F-statistic
should \textsl{not} be a considered a requirement to report empirical
results. The rule-of-thumb that conditional F-statistics should be
at least 10 to signal sufficient instrument strength may be overly
conservative. Instead, weak instrument-robust methods can provide
valid inference in weak and strong instrument settings. We have discussed
three strategies to calculate robust confidence sets with two-sample
summary data based on \citet{Anderson1949}, \citet{Kleibergen2005},
and the linear combination statistic of \citet{Andrews2018}, and
we have provided software to implement their usage. 

When the number of instruments exceed the number of exposures, Anderson-Rubin
confidence sets should be treated with caution if they are much more
precise than Kleibergen and Andrews confidence sets \citep{Davidson2014}.
If Anderson-Rubin confidence sets are empty, then this suggests there
may be excessive heterogeneity in genetic variant associations with
the outcome. In such cases, we may consider the evidence from Kleibergen-OH
confidence sets, which we have designed to account for overdispersion
heterogeneity in genetic variant associations with the outcome. 

Compared with usual summary data Mendelian randomization methods,
the runtime for computing these weak instrument-robust confidence
sets is quite long, but not prohibitively so; taking minutes in a
typical application, rather than seconds. However, the runtime increases
considerably with every additional instrument included, and exponentially
with every additional exposure included. 

Another perceived downside with weak instrument-robust methods is
that confidence sets can be very large, and possibly infinite, under
weak instruments. But as discussed by \citet{Keane2023}, it is perhaps
odd to expect bounded confidence sets when instruments are so weak
that we are not confident the multivariable model being estimated
is even identified. 

\subsection*{Funding acknowledgement}

This research was funded by the United Kingdom Research and Innovation
Medical Research Council (MC\_UU\_00040/01), and supported by the
National Institute for Health Research Cambridge Biomedical Research
Centre (BRC-1215-20014). S.B. is supported by the Wellcome Trust (225790/Z/22/Z). 

\bibliographystyle{chicago}
\bibliography{mvmr_weakIV}

\end{document}